\documentclass[prl,twocolumn,showpacs,amsmath,amssymb]{revtex4}

\usepackage{graphicx}
\usepackage{bm}
\usepackage{color}
\begin{document}

\title{Tracing the thermal mechanism in femtosecond spin dynamics}
\author{U. Atxitia$^{1}$, O. Chubykalo-Fesenko$^{1}$, J. Walowski$^{2}$, A. Mann$^{2}$  and M. M\"{u}nzenberg$^{2}$}

\affiliation{$^{1}$Instituto de Ciencia de Materiales de Madrid,
CSIC, Cantoblanco, 28049 Madrid, Spain}
\affiliation{$^{2}$I. und IV. Physikalisches Institut, Universit\"{a}t G\"{o}ttingen, 37077 G\"{o}ttingen, Germany}

\begin{abstract}
We compare femtosecond pump-probe experiments in Ni and micromagnetic modelling based on the Landau-Lifshitz-Bloch equation coupled to a two-temperature model, revealing a predominant thermal ultrafast demagnetization mechanism. We show that both spin (femtosecond demagnetization) and electron-phonon (magnetization recovery) rates in Ni increase as a function of the laser pump fluence. The slowing down for high fluences arises from the increased longitudinal relaxation time.

\end{abstract}

\pacs{75.40Gb,78.47.+p, 75.70.-i}
\maketitle

The implementation of novel magnetic recording and spintronic devices requires well-funded knowledge of the limits of spin manipulation. Pump-probe experiments with powerful femtosecond lasers \cite{Beaurepaire,Koopmansmode,Koopmans2005, Hansteen, Walowski, DallaLonga, Djordjevic,Carpene} have pushed these limits down to the femtosecond timescale in the past decade. These experiments have attracted many researchers with the aim to understand both fundamental mechanisms of the magnetization dynamics in a strongly out-of-equilibrium regime and to control the magnetic properties of materials on the femtosecond timescale. Very recently the involvement of the spin in thermoelectric processes, spin Peltier or Seebeck effects, has become of strong interest \cite{SpinSeebeck}. Therefore a vital understanding  of energy transport between the spin, electron and phonon system in ferromagnets is needed. Even for the most simple itinerant ferromagnets, such as Ni, the processes connecting the elementary spin scattering process and the terahertz (THz) spin-wave generation have not been identified yet. However, they are the key to understand the macroscopic demagnetization on the femtosecond time scale.

 Different non-thermal mechanisms of how light could couple to the spin system have been put forward: the excitation of non-magnetic states mediated by the enhanced spin-orbit interaction (SOC) \cite{Zhang}, the inverse Faraday \cite{Hansteen} or the Barnet effect \cite{Rebei}. At the moment, no clear proof of these effects has been presented for the ultrafast magnetization dynamics in Ni: (i) The change of the light polarization has no influence on the femtosecond demagnetization and estimations suggest that the amount of direct angular momentum transfer from photons to spins is negligible \cite{DallaLonga}, (ii) the time-dependent density functional theory based on the SOC mechanism at its present state of the art srongly underestimates the experimentally observed timescales in itinerant ferromagnets \cite{Zhang}.

 Differenty from the latter, it has been shown that more applicable is a "thermal" ansatz for the description of the femtosecond magnetization dynamics in the itinerant ferromagnets \cite{Beaurepaire,Hohlfeld00,Koopmans2005, Kazantseva, Kazantseva08}. Within this description, it is assumed that the excited state is a statistical ensemble of many electronic excitations, based on the undisputed fact that the photons of a femtosecond laser, focused on a metal, pass the energy to the subsystems of electrons, phonons and spins. Photons are absorbed by electrons close to the Fermi level leading to a non-equilibrium distribution that thermalizes within several femtoseconds. In the so-called two-temperature ($2T$) model \cite{Kaganov, Schoenlein87, Allen87}, energy dissipation in the system is introduced by including rate equations, artificially separating electrons and lattice by attributing temperatures to both. The stochastic spin-flip processes are driven by the increased electron temperature. The most probable candidate for a microscopic, femtosecond spin-flip mechanism is the phonon- and impurity-mediated Elliott-Yafet scattering \cite{Elliott, Koopmansmode, Walowski, Faehnle}. The corresponding spin-flip dynamics model should be able to directly relate the electronic excitations and the magnetic precessional motion arising from a Landau-Lifshitz type damped precession.  In this line a thermal formalism based on the stochastic Landau-Lifshitz-Gilbert (LLG) equation was used to model the sub-picosecond spin dynamics within an atomistic Heisenberg model by Kazantseva et al. \cite {Kazantseva}. The spin relaxation rates calculated from these models have been found to depend on the strength of the coupling to the electron temperature bath \cite{Kazantseva, Mueller09} and solidified the presence of THz high energy spin waves \cite{Djordjevic, Kazantseva}.

  Different to the discrete Heisenberg model, we use in the follwing a micromagnetic  (\textit{macrospin}) Landau-Lifshitz-Bloch (LLB) equation where the magnetic fluctuations are described by using the temperature and time-dependent parameters. We compare this thermal macrospin model with the measured ultrafast magnetization and reflectivity dynamics in Ni for a set of different excitation powers.  The excellent agreement between experiment and modelling proves that the ultrafast demagnetization in Ni has a purely thermal origin.

Our experiment was performed on a $15\,\mathrm{nm}$ thick ferromagnetic Ni film deposited on a Si(100) substrate using electron beam evaporation in ultra high vacuum ($5\times10^{-10}\,\mathrm{mbar}$). A femtosecond pump-probe experiment was used to determine the reflectivity $R(\tau)$ and the Kerr rotation $\theta_K (\tau)$ as a function of the probe pulse delay $\tau$ \cite{Djordjevic,Koopmansmode} immediately after sample preparation. The exciting pump pulse fluence was varied from $10$ to $50\,\mathrm{mJ/cm^2}$ per pulse ($80\,\mathrm{fs}$, $\lambda=800\,\mathrm{nm}$) calculated from the area  determined from the half width of the Gaussian pump beam intensity profile ($60\,\mathrm{\mu m}$). A static field was applied in-plane to saturate the sample. The Kerr rotation $\theta_K (\tau)$ is defined in the following as the asymmetric part of the signal, $\theta_{K,-}(\tau)={1 \over 2}(\theta_{K}(\tau,M)-\theta_{K}(\tau,-M))$ changing with the field direction and mirroring predominantly the magnetization change. The symmetric part $\theta_{K,+}(\tau)={1 \over 2}(\theta_{K}(\tau,M)+\theta_{K}(\tau,-M))$ mirrors the reflectivity change $R(\tau)$. However the reflectivity change $R(\tau)$ was determined separately by a differential measurement method of the direct sample reflectivity for $p$-polarized probe pulses.

\begin{figure} [h!]
\includegraphics[scale=0.65]{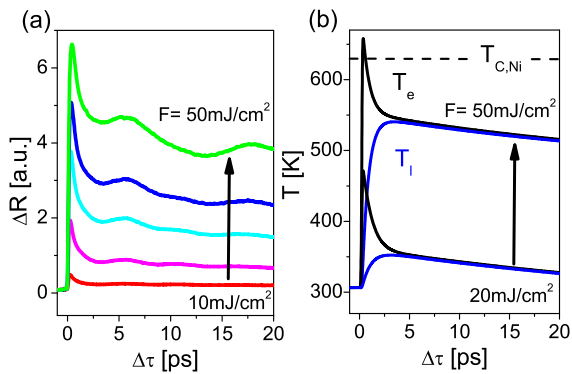}
\caption{(a) Response after femtosecond laser excitation revealing the  dynamics of the heated electrons, coherent stress waves and incoherent lattice excitations at various pump fluences $F$ contributing to the transient reflectivity for a 15 nm Ni thin film. (b) The results of the integration of the $2T$ model for two pump fluence values.}
\label{Refl}
\end{figure}

The data of the delayed reflectivity $R(\tau)$ for various pump fluences are presented in Fig. \ref{Refl}(a) and show oscillations with a main period of about $7\,\mathrm{ps}$. It originates from the coherent excitation of a shockwave (sound velocity $\nu_{Ni}=5400\,\mathrm{m/s}$). In case of a film thickness in the range of $t_{\text{Ni}} \sim  t_{\text{opt},800\,\mathrm{nm}}$ a standing shock wave forms with the wave vector $k=\pi /t_{\text{Ni}}$ \cite{Djordjevic2}.  For a threshold fluence above $30\,\mathrm{mJ/cm^2}$ the strong coherent stress wave excitations appear to have an incoherent counterpart (lattice heating and thermal expansion) owing an effect on the reflectivity.

Neglecting the contribution of the incoherent stress wave excitations, the reflectivity change in the first approximation is defined by the electron temperature $T_e$ \cite{Schoenlein87}. The electron temperature $T_e$ is coupled to the lattice temperature $T_l$ within the $2T$ model \cite{Kaganov} in the form of two differential equations:

\begin{eqnarray}
 C_e {d T_e \over dt} &=& -G_{el} (T_e - T_l) + P(t) \nonumber \\
 C_l {d T_l \over dt} &=& G_{el} (T_e - T_l)-\frac{(T_{l}-T_\text{room})}{\tau_{th}},
\label{temp}
\end{eqnarray}
Here  $C_{e}$ and $C_{l}$ are the specific heats of the electrons and the lattice, $G_{el}$ is an electron-phonon coupling constant which
gives the rate of the energy exchange between the electrons and the
lattice \cite{Allen87} and  $\tau_{\text{th}}$ is the heat diffusion
time. The source term $P(t)$ is a function which
describes the laser power density absorbed in the material. We will assume a Gaussian laser pulse of
$80\,\mathrm{fs}$ duration. The electron specific heat should strongly depend
on $T_{e}$ due to variation of the density of states of Ni around
the Fermi level with temperature. Here we use a simplification for this dependence and assume $C_{e}=\gamma_e T_{e}$, $\gamma_e=3\times10^{3}\,\mathrm{Jm^{-3}K^{-2}}$.
Within this approximation the analysis of the data  in Fig. \ref{Refl}(a), using the method of Ref.\cite{Hohlfeld00}, gives $G_{el}\approx10\times10^{7}\,\mathrm{Wm^{-3}K^{-1}}$
at different
levels of laser excitation, consistent with similar values reported in literature for Ni \cite{Caffrey}.

\begin{figure}[h!]
\includegraphics[scale=0.6]{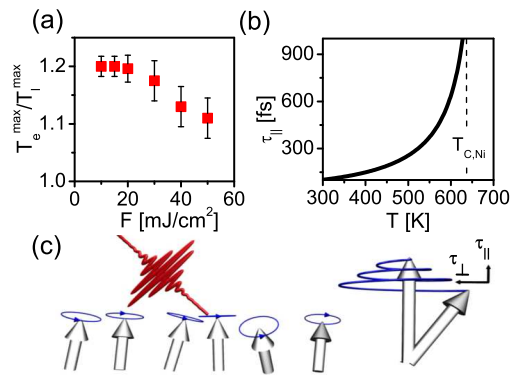}
\caption{Parameters defining the slowing down of the magnetization rates. (a) The ratio between the maximum electron and lattice temperatures $T_{e}^{max}/T_{l}^{max}$ as obtained fitting the $2T$ model to the reflectivity data in Fig. \ref{Refl}(a) as a function of pump fluence. (b) The longitudinal relaxation time for Ni as a function of temperature evaluated from the MFA. (c) Schematics of the thermal spin excitations followed by the femtosecond laser pulse (left) and the thermal macrospin model (right).}
\label{Tle}
\end{figure}

\begin{figure}[h!]
\includegraphics[scale=0.65]{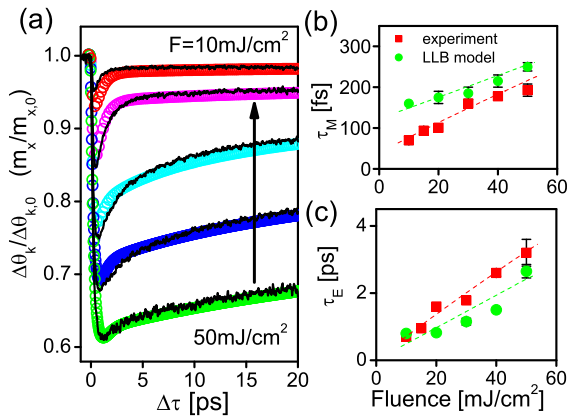}
\caption{(a) Absolute values of the measured dynamics of the Kerr angle rotation (solid lines) and the modeled magnetization dynamics (symbols) for various laser pump
fluences $F$. (Right) The magnetization dynamics rates as a function of pump fluence: (b) demagnetization time $\tau_M$ and (c) magnetization recovery time $\tau_E$ as function of the pump fluence are plotted. Experimental (squares) and modeling (circles) values are given for a direct comparison. Lines are a guide to the eyes. Note that $\tau_E$ differs from the electron-phonon equilibration $\tau_{ep}$ by its magnetic contribution.}
\label{Demag}
\end{figure}

The lattice specific heat at low pump fluences is  taken as $C_{l}=C(\textrm{300\, K})-C_{e}(\textrm{300\, K})=3.1\times10^{6}\,\mathrm{Jm^{-3}K^{-1}}$,
where $C(\textrm{300\, K})=4\times10^{6}\,\mathrm{Jm^{-3}K^{-1}}$ is the experimental total specific heat of Ni \cite{Braun}. The fitting of the $2T$ model to the  reflectivity data from Fig. \ref{Refl}(a) shows that the energy, put into the lattice,  increases as a function of pump fluency, showing a decrease of the $T_{max}^{e}/T_{max}^{l}$ ratio (see Fig. \ref{Tle}b).
This decrease may contain different contributions such as
   (i) nonlinear temperature dependence of the specific heat $C_e$,  (ii) additional energy dissipation mechanisms, e.g. excitation of elastic waves for $T>\Theta_{\text{Debye}}$ ($475\,\mathrm{K}$ in Ni) (iii) dependence of the reflectivity data on $T_l$ at high pump fluencies \cite{Djordjevic2}.  The $2T$ model was adjusted to capture the behavior observed  by decreasing the lattice specific heat from the value above to $C_{l}=1.0\times10^{6}\,\mathrm{Jm^{-3}K^{-1}}$ at high pump fluence.  Finally, the heat diffusion time of $\tau_{th}\approx 50\,\mathrm{ps}$ was obtained by fitting the reflectivity data in Fig. \ref{Refl} for all pump fluences. Figure \ref{Refl}(b) presents the result of the $2T$ model integration (Eqs. (\ref{temp})).  Note that for the highest pump fluence the electron temperature exceeds the Curie temperature $T_{c, Ni}$ while the lattice temperature stays well below that value.

The magnetization dynamics model is based on the LLB micromagnetic equation \cite{Atxitia}.
This macroscopic equation has been derived for thermodynamically averaged spin polarization $\mathbf{m}$  by D. Garanin  within mean field approximation (MFA)
from the classical Fokker-Planck equation for atomistic spins interacting
with a heat bath and from the corresponding density-matrix in the
quantum case \cite{Garanin}. The macrospin LLB equation has been shown to be a valid micromagnetic equation at all temperatures, even above $T_c$ \cite{Chubykalo}. The advantage of the present model is the use of only one thermodynamically consistent macrospin equation, as compared to large-scale calculations based on the atomistic Heisenberg model \cite{Kazantseva}.
We write the LLB equation
as follows:

\begin{eqnarray}
&&\mathbf{\dot{m}}=\gamma \lbrack \mathbf{m}\times  \mathbf{H}_{%
\mathrm{eff}} ]+\frac{\gamma \alpha _{||}}{m^{2}}(%
\mathbf{m}\cdot \left( \mathbf{H}_{\mathrm{eff}}\right) )%
\mathbf{m} \nonumber \\
&&\qquad {}-\frac{\gamma \alpha
_{\perp}}{m^{2}}[\mathbf{m}\times \lbrack
\mathbf{m}\times \left( \mathbf{H}_{\mathrm{eff}}\right) %
]],  \label{LLB}
\end{eqnarray}
where $\gamma$ is the gyromagnetic ratio, $\alpha _{\parallel }$ and
$\alpha _{\perp }$ are dimensionless longitudinal and transverse
damping parameters given by $\alpha_{\parallel}= 2 \lambda T/3T_{c}, \alpha_{\perp}=
\lambda \left[1-T/3T_{c}\right]$ for $T<T_c$ and $\alpha _{\perp }=
\alpha _{\parallel }$ for $T>T_c$.  Here $\lambda$ is the parameter
describing the coupling of the spins to the heat bath, which is assumed to be the electron system for the present case, as suggested in Ref. \cite{Kazantseva}.  The effective field $\mathbf{H}_{\mathrm{eff}}$ is given by

\begin{equation}
 \mathbf{H}_{\mathrm{eff}}=\mathbf{H}+\mathbf{H}_{A}+\left\{
 \begin{array}{cc}
   \frac{1}{2\tilde{\chi}_{\Vert }}\left(
   1-\frac{m^{2}}{m_{e}^{2}}\right) \mathbf{m} & T\lesssim T_{c} \\
   -\frac{1}{\tilde{\chi}_{\Vert }}\left(1-\frac{3T_c}{5(T- T_c)}m^{2}\right)
   \mathbf{m} & T\gtrsim T_{c}
 \end{array}
 \right. \mathbf{.}  \label{Heffm}
\end{equation}
In our model the temperature-dependent equilibrium magnetization $m_e$ and longitudinal susceptibility $\tilde{\chi}_{\Vert }$ were determined in MFA \cite{Atxitia}, adjusting the Curie temperature value to the experimental one $T_{c}=631\,\mathrm{K}$. In a more rigorous multiscale description \cite{Kazantseva08}, they could be evaluated directly from the spin Hamiltonian, parameterized from the electronic structure calculations, or considered as an experimental input. We use the following parameters for Ni: The saturation
magnetization $M_{s}=500\mathrm{emu/cm^3}$ (at $T=0\,\mathrm{K}$), the applied field $\mathbf{H}=1500\,\mathbf{e}_{x}\,\mathrm{Oe}$, the damping parameter $\alpha_{\bot}=0.04$ at $T=300\,\mathrm{K}$ \cite{Djordjevic}
(from which the intrinsic coupling parameter $\lambda=0.045$ is estimated). The easy plane anisotropy, defined by magnetostatic interactions for the thin film geometry, is given by $\mathbf{H}_{A}=4\pi M_{s}\left(m_{x}\mathbf{e}_{x}+m_{y}\mathbf{e}_{y}\right)$. The magneto-crystalline anisotropy is neglected.

One of the immediate consequences of the thermal dynamics based on the LLB equation is the presence of the longitudinal relaxation time, i.e. the rate at which the spin system  disorders at a given temperature.
The longitudinal relaxation time mirrors the appearance of short wavelength spin exciations at THz frequency in the thermal microscopic model.
In the absence of the anisotropy the longitudinal relaxation time behaves as $\tau_{||}(T)=\tilde{\chi}_{||}(T,\textbf{H})/\gamma\alpha_{||}(T)$ where $\tilde{\chi}_{||}(T,\textbf{H})$ is the longitudinal susceptibility at applied field \textbf{H} \cite{Chubykalo}. The dependence of the longitudinal relaxation time on the temperature for Ni is presented in Fig. \ref{Tle}(b).
It strongly increases at $T_{c, Ni}$, due to a faster divergence of the longitudinal susceptibility as compared to the variation of the longitudinal damping; a general critical phenomenon known as the critical slowing down \cite{Chen}.
 It is in the order of $0.1\,\mathrm{ps}$ for low temperatures and increases by more than one order of magnitude towards $T_c$.

 Therefrom one concludes that during the laser-induced demagnetization, the electron temperature is changed on the timescale faster than the longitudinal relaxation time. Consequently,  the spin system cannot follow the electron temperature. The magnetic response is delayed by the slowing down of the longitudinal relaxation being inversely proportional to the longitudinal damping constant $\alpha_{||}$. Furthermore, $\alpha_{||}(T)$ is related to the Gilbert damping $\alpha_{\perp}(T)$  via the coupling parameter $\lambda$. Therefrom one derives a temperature-dependent correction to the model of B. Koopmans \textit{et al.} \cite{Koopmans2005}, relating the femtosecond demagnetization time to the Gilbert damping parameter by $\tau_M \sim 1/\alpha_{\perp}$.

 For a quantitative comparison the results of the integration of the system of Eqs. (\ref{temp}) and (\ref{LLB}) are presented in Fig. \ref{Demag} (a) in comparison to the experimental time-resolved Kerr angle rotation data. Both show the slowing down of the time scales as the pump fluence is increased from $10\,\mathrm{mJ/cm^2}$ to $50\,\mathrm{mJ/cm^2}$. Fig. \ref{Demag}(a) shows a very good agreement between theory and experiment, giving the absolute value of the demagnetization as 3\% for the lowest pump fluence to around 40\% for the highest. To accurately determine the relaxation times, the slopes are analyzed using the analytic solution for the three-temperature-model (including the spin temperature) given by Dalla Longa \textit{et al.} \cite{DallaLonga}:

\begin{eqnarray}
-\frac{\Delta M_{x}(t)}{M_{0,x}}=   \Big\{   \Big[A_{1}F(\tau_{0},t)-\frac{(A_{2}\tau_{E}-A_{1}\tau_{M})e^{-t/\tau_{M}}}{\tau_{E}-\tau_{M}}\nonumber \\
-\frac{\tau_{E}(A_{1}-A_{2})e^{-t/\tau_{E}}}{\tau_{E}-\tau_{M}}\Big]\Theta(t)\Big\} \ast G(t)
\end{eqnarray}
Here $\ast G(t)$ represents the convolution product with the Gaussian laser pulse profile, $\Theta(t)$ is the step function, $A_i$ are the fitting constants and $\tau_M$ and $\tau_E$ define the scale of the magnetization loss and magnetization recovery time, respectively.
The results are presented in Fig. \ref{Demag}(b and c). We observe a slowing of both magnetization rates $\tau_M$ and $\tau_E$ as a function of laser pump fluence, following the slowing down of the longitudinal relaxation time. As a result, the electron-phonon equilibration time extracted from the reflectivity ($\tau_{ep}$) and from the magnetization recovery time ($\tau_{E}$) have different magnitudes. This observation was also reported for Fe \cite{Carpene}. However, for the slowing down of the magnetization recovery, the longitudinal relaxation slowing down is not sufficient itself. In Ref. \cite{Kazantseva} the experimentally observed slowing down of $\tau_E$ in CoPt was attributed to the loss of the magnetization correlations for high pump fluences. In our experiment the combined measurements of static and dynamic Kerr effect have allowed us to determine that the demagnetization for the highest pump fluency was around 40\%, not sufficient to produce the magnetization correlation loss. Our detailed analysis of the reflectivity data reveals that at high pump fluences, the relative energy loss is faster in the electron system than in lattice, mirrored by the ratio between maximum electron and lattice temperatures presented in Fig. \ref{Tle}(a): The cooling dynamics of the lattice and electrons, together with the longitudinal relaxation, leads to the observed slowing of the magnetization recovery. To summarize the result of the direct comparison, the thermal macrospin model, based on the concept of a longitudinal relaxation time in combination with the $2T$ model, fully describes the absolute variation and characteristic time scales of the femtosecond demagnetization experiment in Ni.

 In conclusion, the agreement between the thermal macrospin model and the experimental data reveals the dominant character of the thermal demagnetization mechanism in Ni mediated by hot electrons. This thermal magnetization dynamics model has been used to quantitatively describe the absolute values and characteristic timescales determined in the ultrafast magnetization and reflectivity data with increasing the excitation fluence. As a result the theoretical model allows us to identify the longitudinal relaxation as a limiting factor for the slowing down of the demagnetization rates, a general phenomenon which should be present in all ferromagnetic materials. Similarly, other 3d-metals such as Fe, where the high temperature dynamics of the itinerant electrons is suitable for the heat bath description, can be described. However, at the same time the longitudinal relaxation rate is also dependent on the coupling strength between electron and spin. The latter means that the magnitude of the demagnetization rate strongly depends on the microscopic electron, lattice and spin coupling mechanisms. This suggests that femtosecond pump-probe experiments are a valuable tool to characterize spin based caloric effects of a particular material.

 This work was supported by the Spanish projects MAT2007-66719-C03-01, CS2008-023, S-0505/MAT/0194 and the German Research Foundation within SPP 1133.


\end{document}